\documentclass[
prl,
superscriptaddress,amsmath,amssymb,
prb,
twocolumn,
]{revtex4-1}

\usepackage{graphicx}
\DeclareGraphicsExtensions{.pdf,.eps,.png,.jpg}
\usepackage{dcolumn}
\usepackage{float}
\usepackage{bm}
\usepackage{xspace}
\usepackage{color}
\newcommand{\cvs}{CsV$_3$Sb$_5$\xspace}
\newcommand{\thigh}{T$_\mathrm{c1}$\xspace}
\newcommand{\tlow}{T$_\mathrm{c2}$\xspace}
\newcommand{\sg}{$P6/mmm$\xspace}
\newcommand{\half}{\mbox{(0.5,0.5,0.5)}\xspace}
\newcommand{\quadr}{\mbox{(0.5,0.5,0.25)}\xspace}
\newcommand{\RN}[1]{\uppercase\expandafter{\romannumeral#1}}

\begin{document}

\title{Temperature-driven reorganization of electronic order in \cvs}

\author{Q. Stahl}
\affiliation{Institute of Solid State and Materials Physics,  TU Dresden, 01069 Dresden, Germany}
\author{D. Chen}
\affiliation{Max Planck Institute for Chemical Physics of Solids, Dresden, Germany}
\author{T. Ritschel}
\affiliation{Institute of Solid State and Materials Physics,  TU Dresden, 01069 Dresden, Germany}
\author{C. Shekhar}
\affiliation{Max Planck Institute for Chemical Physics of Solids, Dresden, Germany}
\author{C. Felser}
\affiliation{Max Planck Institute for Chemical Physics of Solids, Dresden, Germany}
\affiliation{W\"urzburg-Dresden Cluster of Excellence ct.qmat, Technische Universit\"at Dresden, 01062 Dresden, Germany}
\author{J. Geck}
\email[]{jochen.geck@tu-dresden.de}
\affiliation{Institute of Solid State and Materials Physics,  TU Dresden, 01069 Dresden, Germany}
\affiliation{W\"urzburg-Dresden Cluster of Excellence ct.qmat, Technische Universit\"at Dresden, 01062 Dresden, Germany}


\date{\today}


\begin{abstract}
We report a temperature dependent x-ray diffraction study of the electronic ordering instabilities in the kagome material \cvs. Our measurements between 10\,K and 120\,K reveal an unexpected reorganization of the three-dimensional electronic order in the bulk of \cvs: At 10\,K, a 2x2x2 superstructure modulation due to electronic order is observed, which upon warming changes to a 2x2x4 superstructure at 60\,K. The electronic order-order transition discovered here involves a change in the stacking of electronically ordered V$_3$Sb$_5$-layers and agrees perfectly with anomalies previously observed in magneto-transport measurements. This implies that the temperature dependent three-dimensional electronic order plays a decisive role for transport properties, which are related to the Berry-curvature of the V-bands. Our data also show that the bulk electronic order in \cvs breaks the 6-fold rotational symmetry of the underlying \sg lattice structure.
\end{abstract}

\maketitle

The physics of quantum materials containing layers of corner-sharing triangles is full of surprises. Not only can coupled magnetic moments on such a kagome lattice exhibit unconventional magnetic ground states\,\cite{Sachdev:1992d}. As it is now becoming more and more evident, electrons on a kagome lattice also provide a truly fascinating platform for various other collective quantum phenomena, such as nontrivial topological matter\,\cite{Guo:2009v,Bilitewski:2018z}, chiral superconductivity\,\cite{Yu:2012y, Kiesel:2012a}, bond and charge ordering\,\cite{Wang:2013p, Kiesel:2013p} or complex intertwined order\,\cite{Yu:2021v, Zhao:2021a}.

In this regard the kagome materials $A$V$_3$Sb$_5$ ($A=$K, Rb, Cs) have become particularly famous, generating a real surge of research into their electronic properties\,\cite{Ortiz:2020h,denner2021,Ortiz:2021a,Jiang:2021z, Wang:2021a, Zhao:2021a,Chen:2021a,Liang:2021a}. One reason for the large interest certainly is the appearance of superconductivity in a topological $Z_2$ kagome metal\,\cite{Ortiz:2020h}. Superconductivity in layered kagome materials is very rare to start with. But of course the coexistence of superconductivity and non-trivial electronic topology may support elusive and highly sought after quasiparticles like Majorana bound states\,\cite{Liang:2021a}.  
In addition to this, intriguing electronic ordering instabilities have been observed in $A$V$_3$Sb$_5$. Very recent scanning tunneling microscopy (STM) studies revealed the presence of electronic orders on the surfaces of  KV$_3$Sb$_5$ and \cvs\cite{Jiang:2021z, Wang:2021a, Zhao:2021a}.

\begin{figure}[b!]
    \centering
    \includegraphics[width=0.85\columnwidth]{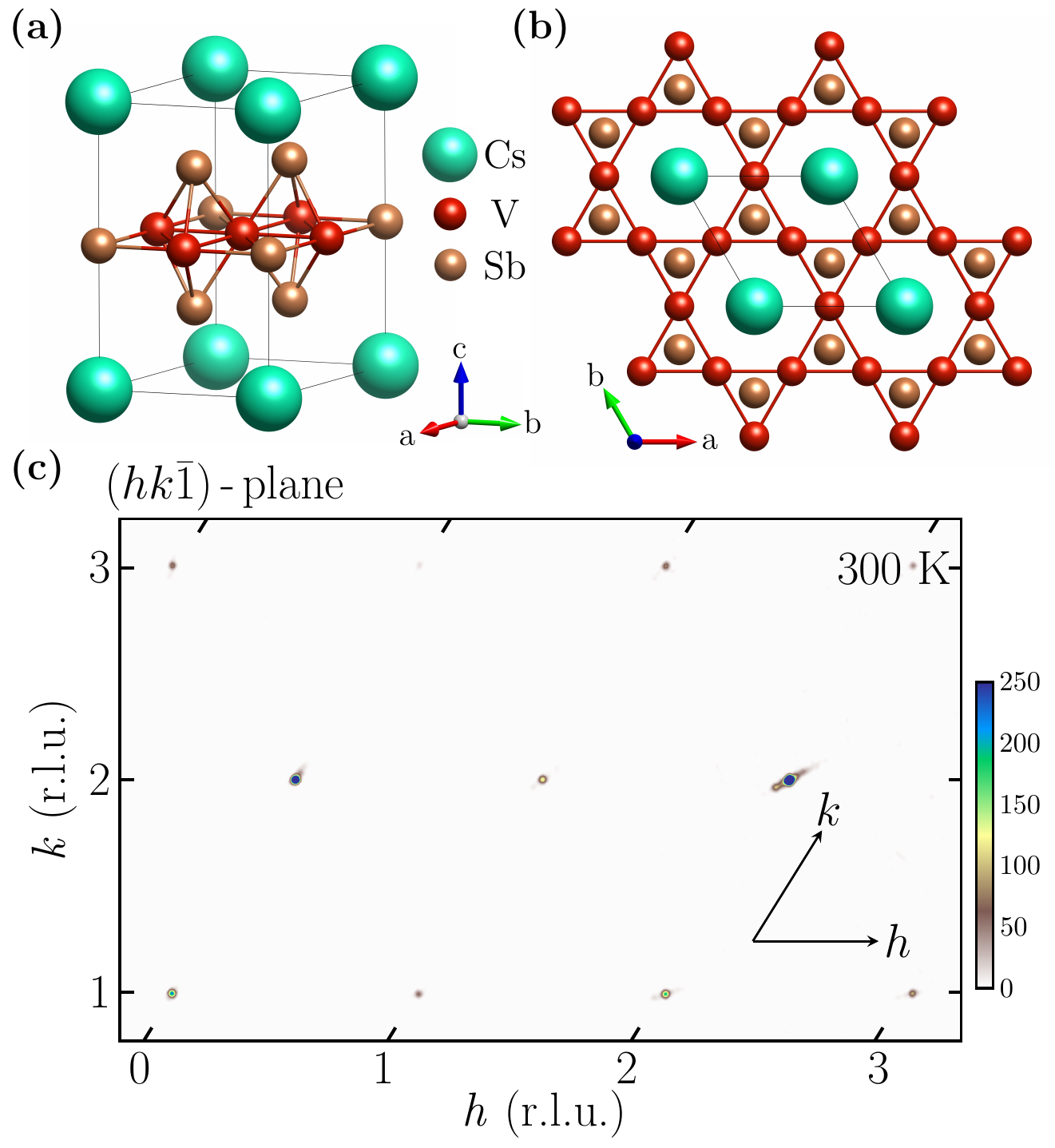}
    \caption{(a), (b): Illustration of the three-dimensional room temperature \sg  structure of \cvs (left) and V$_3$Sb$_5$-layer with its a kagome sublattice of V-sites (right). (c): Representative distribution of the XRD-intensity in the ($hk\bar1$)\,-\,plane measured at room temperature. The color scale highlights the low intensities up to 250\,cps. The highest intensity at (2,2,-1) is of the order of $10^4$\,cps. }
    \label{fig:rt_structure}
\end{figure}

Focusing on the case of \cvs, electronic order sets in below 95\,K \cite{Ortiz:2020h,Wang:2021a}, resulting a 2x2 superstructure, which in fact shows signatures of chirality\,\cite{Wang:2021a}. 
The latter are indeed in perfect agreement with previous theoretical studies of the kagome Hubbard-model where such an instability has been predicted\,\cite{Kiesel:2012a, Kiesel:2013p}. 
Importantly, the electronic order in \cvs is three-dimensional (3D), i.e. there are also correlations between the kagome layers along the c-axis (c.f. Fig.\,\ref{fig:rt_structure})\,\cite{Liang:2021a, Ortiz:2021a, ratcliff:2021condmat}.
These correlations along $c$ remain, however, controversial: while some studies find clear indications for a 2x2x2 order\,\cite{Liang:2021a, ratcliff:2021condmat}, a very recent x-ray diffraction (XRD) study revealed a disordered  2x2x4 modulation\,\cite{Ortiz:2021a}. Also, it remains an open question as to whether these correlations along $c$ play any significant role at all for the electronic properties of \cvs.

\begin{figure*}[t]
    \centering
    \includegraphics[width=0.85\textwidth]{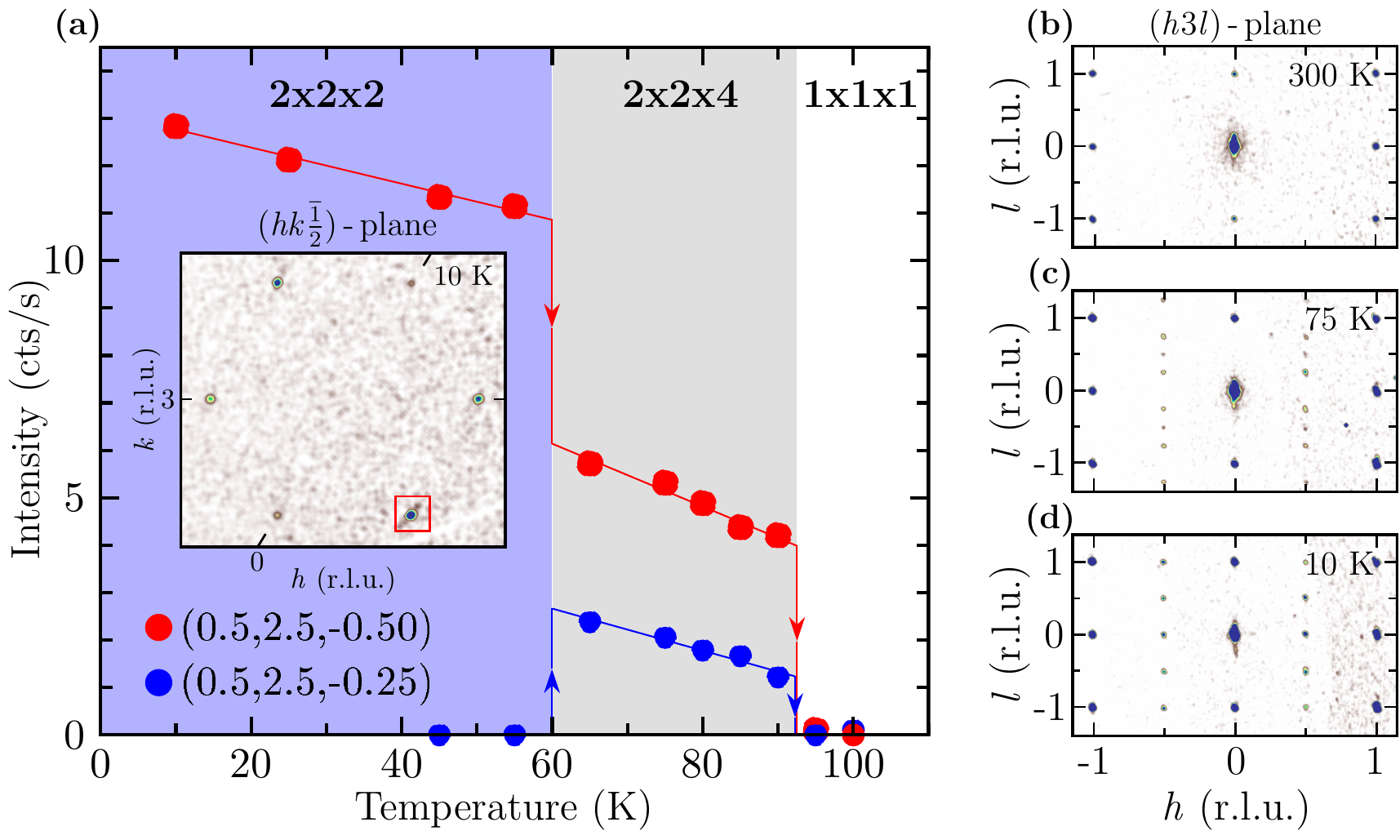}
    \caption{Temperature dependent XRD-data for \cvs, revealing the appearance of two distinct 3D superlattice modulations with different out-of-plane correlations along $c$. Data were taken with increasing temperature. (a) Temperature dependent integrated intensity of the (0.5,2.5,-0.5) and (0.5,2.5,-0.25) superstructure reflections, which signal the presence of a 2x2x2 and a 2x2x4 superlattice, respectively. The red and blue lines serve as guides to the eye. Inset: Reconstructed reciprocal space map parallel to the ($hk\bar{\frac{1}{2}}$)\,-\,plane for the 2x2x2 phase at 10 K. The chirality of the XRD-intensity is similar to that observed by STM\,\cite{Wang:2021a}. 
    $(h3l)$\,-\,planes for the undistorted 1x1x1 host phase (b) at 300\,K, the 2x2x4 phase (c) at 75\,K and the 2x2x2 phase (d) at 10\,K. The indexation always corresponds to the undistorted 1x1x1 host phase (\sg).
}
    \label{fig:overview}
\end{figure*}

Here we study the 3D electronic order in the bulk of \cvs single crystals using XRD. We show that, besides the electronic order-disorder transition at about 95\,K, there is an additional order-order transition at 60\,K. This transition agrees perfectly with previously reported anomalies in the magneto-transport\,\cite{Chen:2021a, Xiang:2021a}, implying that these changes in the electronic order play a decisive role for the topological band structure of \cvs. 

\begin{figure*}[t]
    \centering
    \includegraphics[width=0.85\textwidth]{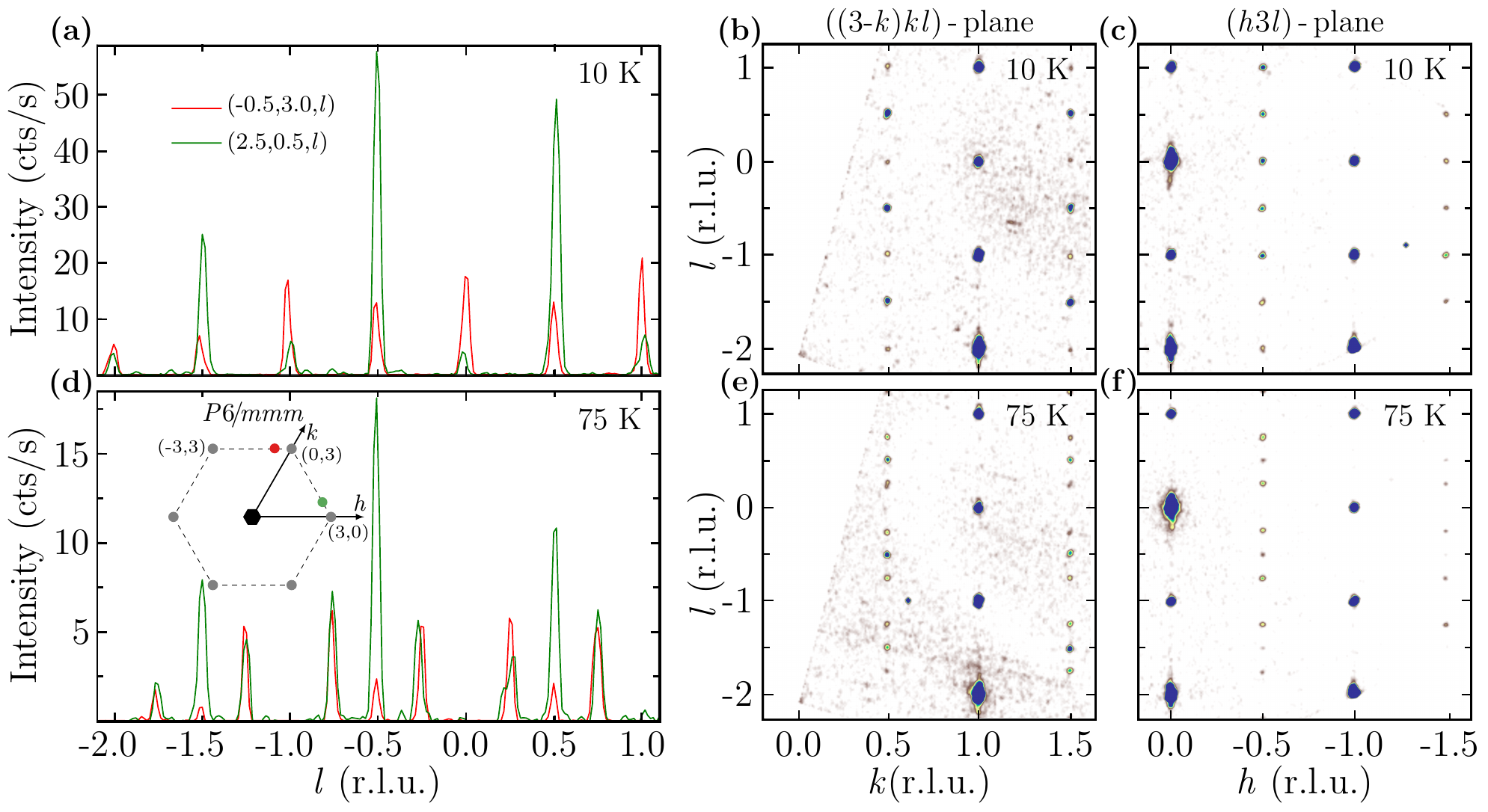}
    \caption{
    Breaking of the 6-fold rotational symmetry due to electronic order in the 2x2x2 and 2x2x4 phase. (a), (d): $l$-cuts  $(-0.5,3.0,l)$ and $(2.5,0.5,l)$, which are related by 6-fold symmetry, differ in both phases (integration thickness $\delta h$ = $\delta k$ = 0.06 r.l.u.). Also the intensity distributions in reciprocal lattice planes related by 6-fold symmetry are distinct as demonstrated in (b), (c) (2x2x2 phase) and (e), (f) (2x2x4). Inset: Schematic, indicating the $hk$-positions of the measurements in (a),(d). The indexation corresponds to the undistorted 1x1x1 host phase (\sg). The integration time for the measurements in (a) and (d) was 2160\,s and 2880\,s, respectively. Integration time for (b),(c) was 270\,s and for (e),(f) 360\,s.
    }
    \label{fig:symmetry}
\end{figure*}

We investigated the bulk electronic order in \cvs, using a high-performance laboratory facility for XRD optimized for resolution and sensitivity. This custom-made instrument is equipped with a monochromatized Mo K$_{\beta}$ radiation source (beam diameter $70\mu\mathrm{m}$ on the sample) and a 300K CdTe area detector with no read-out noise for high detection efficiency and minimum background. The outstanding performance of the instrument is particularly well demonstrated in Fig.\,\ref{fig:symmetry}\,(d) where peak profiles with a maximum intensity of 2 count/sec are still fully resolved.
The cooling of the sample down to 10\,K was accomplished by a low-vibration pulse-tube cryostat, which was itself mounted on a specialized four-circle diffractometer.

The single crystals for the present XRD study were synthesized using the self-flux method and well characterized prior to the XRD-experiments.  
The magneto-thermal transport measurements reveal the high-quality of our \cvs single crystals and show a large anomalous Nernst effect. Both the Nernst and Seebeck signals exhibit quantum oscillations, which start around 2\,T\,\cite{chen2021c}.
Fig.\,\ref{fig:rt_structure}\,(c) shows a representative diffraction pattern of the \cvs single crystal inside the cryostat at room temperature. The very sharp and resolution-limited diffraction spots (typical FWHM=0.04\,r.l.u.) show the extremely high quality of the studied single crystals. The diffraction pattern could be indexed by a hexagonal unit cell with lattice parameter a = b = 5.5159\,\AA~ and c = 9.2767\,\AA. In total, 145 Bragg reflections could be reached through the windows of the cryostat, 47 of them being unique. Refinement of the measured intensities\,\cite{note1} yielded the expected \sg structure and $z_{Sb2}$=0.742 as single atomic coordinate not fixed by a special position in perfect agreement with previously published data\,\cite{Ortiz2019}.

Starting from room temperature, the sample was cooled down very carefully to 10\,K at a rate of about 2\,K per minute. After cooling down, a large number of additional superlattice reflections was observed at 10\,K. The inset of Fig.\,\ref{fig:overview}\,(a), Fig.\,\ref{fig:overview}\,(d) and  Fig.\,\ref{fig:symmetry}\,(a)--(c) show representative reciprocal space maps and scans with typical superlattice peaks. 
%
%
The superlattice reflections at 10\,K correspond to a modulation with  wave vector \half, which means that the lattice period is doubled along $a$, $b$ and $c$.
We will refer to this as 2x2x2 superlattice in the following. The superlattice reflections are extremely narrow and resolution-limited (FWHM $\simeq0.04$\,r.l.u.) in all three directions of reciprocal space, meaning that the structural modulation probed by XRD is long-ranged in all three directions of direct space, i.e. we are dealing with a fully developed 3D-order.

We note here that the electronic order at the surface is unstable and  transforms into a stripe-like $4\times1$ surface modulation at low temperatures\,\cite{Zhao:2021a}. The earlier attribution of this instability to surface effects\,\cite{Wang:2021a} is indeed supported by the absence of corresponding superlattice peaks in our bulk sensitive XRD.

The main result of the present study concerns the temperature evolution of the superlattice, which is displayed in Fig.\,\ref{fig:overview}\,(a): Upon warming, the average intensity at the $(0.5,0.5,0.5)$-positions decreases continuously up to about 55\,K and then drops abruptly at about 60\,K. Just above 60\,K we observe additional peaks at positions revealing a quadrupling along the $c$-direction of the \sg unit cell. This reveals the formation of a modulation with wave vector \quadr, i.e. a 2x2x4 superlattice. Note that only the period of the modulation along $c$ is changed, while the periodicity of the superlattice along $a$ and $b$ remains unaltered.
Also the \quadr superlattice peaks are resolution limited, which can be observed in Figs\,\ref{fig:overview}\,(b) and \ref{fig:symmetry}\,(d),(e) and (f). The 2x2x4 modulation therefore as well is long-ranged ordered in all three directions of space. With further increasing temperature, the decrease of the peak intensities continues, followed by an abrupt drop to zero intensity at about 95\,K.

We point out that our data excludes a coexistence of the 2x2x2 and 2x2x4 phase at the measured temperatures between 60\,K and 95\,K: The intensity of superlattice peaks with $l=0$ is always finite in the 2x2x2 phase, whereas it vanishes completely in the 2x2x4 phase, which is incompatible with a phase coexistence. In addition, the intensities of superlattice peaks with $l=0.5$ and $l=0.25$ above 60\,K exhibit a very similar temperature dependence, as can be seen in Fig.\,\ref{fig:overview}\,(a), further supporting the conclusion that all superlattice peaks have the same origin, namely the 2x2x4 phase.


The doubling along the $a$- and $b$-direction observed in our XRD measurements is in perfect agreement with the 2x2 superstructures observed at the surface by STM. This surface ordering therefore corresponds to the ordering in the bulk. The agreement between STM and XRD also provides compelling evidence for the direct relation between the structural modulations detected by XRD and electronic order in the V$_3$Sb$_5$-layers probed by STM.
We therefore conclude that also the structural transition at 60\,K  reflects a change in the electronic order. More specifically, the jump $(0.5,0.5,0.5)\rightarrow(0.5,0.5,0,25)$ of the modulation vector reveals a reorganization of electronically ordered kagome layers along $c$, while the period of the electronic order in the $ab$-plane remains unchanged. 

We note that a previous XRD study reported a 2x2x4 phase at 15\,K\,\cite{Ortiz:2021a}, instead of the  2x2x2 phase observed here. However, the XRD data in Ref.\,\onlinecite{Ortiz:2021a} displays broad streaks along the reciprocal $l$-direction, indicating a considerable amount of disorder. While we cannot provide a definite explanation for this discrepancy, it might very well be that the phase reported in Ref.\,\onlinecite{Ortiz:2021a} is a supercooled \quadr-phase. Of course, this needs to be scrutinized further in future studies. 

We argue that the present results show the temperature-dependence of the electronic order in thermal equilibrium for several reasons: Firstly, the superlattices observed here are highly ordered in all three dimensions, which is a very strong indication that these structures belong to the thermodynamically stable state. Secondly, the observed transition temperatures agree perfectly with a number of previous reports:
First of all, the transition at \thigh=95\,K is fully consistent with anomalies observed in the magnetic susceptibility, electrical resistivity and the magnetic heat\,\cite{Ortiz:2020h, Wang:2021a}. But more importantly, the transition at \tlow=60\,K is in perfect agreement with previous magneto-transport studies where precisely at this temperature striking anomalies were observed\,\cite{Chen:2021a, Xiang:2021a}. It also corresponds very closely to drastic change in the temperature dependent Seebeck coefficient\,\cite{chen2021c}. The rearrangement of the electronic order at \tlow is therefore directly related to changes in the macroscopic properties.

This raises the important question as to what the microscopic structure of the electronic order really is. Unfortunately, possible twinning of the superstructure in our sample currently prevents a full refinement of the superlattice modulation. There are, however, a few important conclusions that can still be drawn: In Figs.\,\ref{fig:symmetry}\,(a),(d)  $l$-cuts at two $hk$-positions, which are equivalent in \sg, are shown. Clearly, the $l$-cuts at the two positions differ, showing that the 6-fold symmetry of \sg is indeed broken by the electronic order in both the 2x2x2 and the 2x2x4 phase. This can also be observed very nicely in panels (b),(c) and (e),(f), which compare the intensity distributions in reciprocal lattice planes related by the 6-fold symmetry. Our data therefore excludes high-symmetry charge ordering models with \sg symmetry discussed in the literature\,\cite{christensen2021theory}.
However, our data are compatible lower-symmetry models discussed in the literature, such as star of David or inverse star of David orders as well as a combined stacking of the two. 

A very intriguing feature of the data shown in Fig.\,\ref{fig:symmetry}\,(d) concerns the symmetry of superlattice peaks with $l=0.5$ and $l=0.25$. We indeed find that the peak intensities for $l=0.25$ still possess the 6-fold rotation symmetry, while the ones with $l=0.5$ do not. In other words, part of the superlattice-intensity remains six fold symmetric in the 2x2x4 phase, whereas this symmetry is fully lost in the 2x2x2 phase. In this sense the breaking of the six-fold symmetry gets stronger and stronger upon cooling. 
This observation agrees very nicely with scenarios explaining the magneto-transport anomalies found around 60\,K in terms of rotational symmetry breaking and a possible Pomeranchuk instability\,\cite{Zhao:2021a}.

Now we turn to another very striking feature of the XRD-intensity distribution, namely its chirality. As can be observed in the inset of Fig\,\ref{fig:overview}\,(a), the intensities exhibit a chirality in the $(hk)$-plane very similar to what has been observed by STM\,\cite{Jiang:2021z,Wang:2021a}. Indeed, chiral electronic order yielding time-reversal symmetry is found in the kagome Hubbard model, where the chirality within a single kagome plane results from sublattice interference\,\cite{Kiesel:2012a,Kiesel:2013p}. Interestingly, a very recent $\mu$SR-study of KV$_3$Sb$_5$ reports the observation of time-reversal symmetry breaking due to electronic order\,\cite{mielke2021c}. These $\mu$SR-data also show additional anomalies at lower temperature, which may correspond to the transition at \tlow=60\,K.
%

But when interpreting the chiral XRD-intensity, care must be taken: Twinning of a non-chiral superstructure, for example, can also yield chiral XRD-intensity distributions. The chirality of the XRD-intensities presented in the inset of Fig.\,\ref{fig:overview}\,(a) can therefore not be taken as proof for chiral electronic order. Notwithstanding, the present XRD-results are fully consistent with chiral electronic order in \cvs. 

To conclude, the present study provides new and valuable experimental insight regarding the relation between electronic order, topological band structure and macroscopic transport in the correlated kagome metal \cvs. But the full microscopic architecture of the electronic order and the role played by translational and rotational symmetry breaking remains to be fully understood. This certainly deserves further study and will very likely yield more surprises in the future.


\section*{Acknowledgements}
This research has been supported by the Deutsche Forschungsgemeinschaft through the the projects C06 and C09 of the SFB 1143 (project-id 247310070) and the W\"urzburg-Dresden Cluster of Excellence on Complexity and Topology in Quantum Matter–ct.qmat (EXC 2147, project-id 390858490). CF, DC and CS also gratefully acknowledge additional support by the European Research Council Advanced Grant (No. 742068) “TOPMAT” and the Deutsche Forschungsgemeinschaft (project-id No. 258499086).


%

\end{document}